\documentclass[aps,prd,onecolumn,preprintnumbers,groupedaddress,showpacs,nofootinbib,amssymb]{revtex4}
\usepackage[T1]{fontenc}
\usepackage[latin1]{inputenc}
\usepackage{graphicx}
\usepackage[english]{babel}
\usepackage{amsmath}
\usepackage{amssymb}
\usepackage{amsfonts}
\begin{document}
\def\pp{{\, \mid \hskip -1.5mm =}}
\def\cL{{\cal L}}
\def\be{\begin{equation}}
\def\ee{\end{equation}}
\def\bea{\begin{eqnarray}}
\def\eea{\end{eqnarray}}
\def\tr{{\rm tr}\, }
\def\nn{\nonumber \\}
\def\e{\mathrm{e}}
\def\dac{\displaystyle\frac}
\def\d{\partial}
\def\({\left(}
\def\){\right)}
\def\p{\phantom}
\def\L{\mathcal{L}}
\newcommand{\arcth}{\mathop{\rm arcth}\nolimits}
\newcommand{\diag}{\mathop{\rm diag}\nolimits}
\newcommand{\const}{\mathop{\rm const}\nolimits}

\title{Higher-Order Gauss-Bonnet  Cosmology}

\author{Salvatore Capozziello$^{1,2}$, Mauro Francaviglia$^{3,4}$ and Andrey N. Makarenko $^5$ 
}

\affiliation{$^1$Dipartimento di Fisica, Universit\`{a} di Napoli "Federico II", Napoli, Italy\\
$^2$INFN Sez. di Napoli, Compl. Univ. di Monte S. Angelo, Edificio G, Via Cinthia, I-80126, Napoli, Italy,\\
$^3$Dipartimento di Matematica, Universit\`{a} di Torino, Via Carlo Alberto 10,  and $^4$INFN Sez. di Torino, I-10123 Torino,  Italy,\\
$^5$ Tomsk State Pedagogical University, Tomsk, Russia \\
}

\pacs{04.50.Kd, 04.20.Jb, 04.20.Cv, 98.80.Jk}

\begin{abstract}
We study   cosmological models derived from   higher-order Gauss-Bonnet  gravity  $F(R,G)$ by using  the Lagrange multiplier approach  without assuming the presence of additional fields with the exception of standard perfect fluid  matter.  The presence of  Lagrange multipliers  reduces the number of allowed  solutions. We need to introduce  compatibility conditions  of the FRW equations, which impose strict restrictions on the metric or require the introduction of additional exotic matter. Several classes of $F(R,G)$ models are generated and discussed.
\end{abstract}

\maketitle

\section{Introduction}

Astrophysical data indicate that the observed universe is  in an
accelerated phase \cite{Dat}. This acceleration could be  induced
by  the so-called dark energy
(see Ref. \cite{review} for a recent review and references therein) the nature and properties of which are
not yet understood at fundamental level.

There are several models for dark energy, but, in general, such a constituent can be    figured out, in a coarse grain approach,   as being
constituted  by
some ideal fluid whose equation of state (EoS) 
exhibits  non-standard  properties, in particular, a non  standard adiabatic index.
On the other hand, dark energy can  be considered as a global phenomenon associated with modifications of  
gravity  \cite{review1}, which  are given by extensions 
of General Relativity (implying modifications of the Hilbert--Einstein action
by introducing  scalar fields, curvature invariants like the Ricci scalar $R$ or the Ricci and Riemann tensors  $R_{\mu\nu}$ and $R^{\alpha}_{\beta\mu\nu}$ or the  Gauss--Bonnet topological invariant  \cite{GB11,Odin2}). Such extended actions describe  effective theories coming from fundamental interactions   \cite{faraoni} and, from a cosmological point of view, they should consistently
describe the early-time inflation  and late-time acceleration,
without the introduction of any other material dark component. In this picture, the dark side of the universe could be traced as the lack of a  final theory of gravity where gravitational interaction does not behave in the same way at all scales \cite{silvio}.

In fact, General Relativity presents problems at ultra-violet  (Quantum Gravity) and infra-red  (Cosmology) scales \cite{faraoni,annalen}. 
A similar situation appears for dark matter phenomena. No fundamental candidate has been revealed up to now and dynamics of self-gravitating structures could be addressed by modifications of gravity like $f(R)$-gravity (see for example \cite{annalen1}).
A further problem is  that  it is not clear why  dark energy had no effect at  early 
epochs while it gives dominant contributions in  today observed  universe. According to the 
latest observational data, dark energy currently accounts for about $68\div70$ \% of the total mass-energy amount of
the universe (see, for example, the very recent  data coming from the PLANCK collaboration \cite{Kowalski}).
The main feature of dark energy is that its EoS parameter  $w_D$ is
negative:
\begin{equation}
w_D=p_D/\rho_D<-1/3, 
\end{equation} where
$\rho_D$ is the dark energy density and $p_D$ the pressure. Due to this feature, the Hubble fluid is accelerating instead of decelerating as it should be if governed by ordinary perfect fluid  matter.
Although current data favor the standard $\Lambda$CDM cosmology,
the uncertainties in the determination of the EoS dark energy parameter $w$ are
still too large, namely $w=-1.04^{+0.09}_{-0.10}$. Hence, one is not able to 
determine,
without doubt, which of the three cases:
$w < -1$, $w = -1$, or $w >-1$ is the one actually realized in our universe
\cite{PDP,Amman}.  Furthermore, the evolution of the cosmological constant is a very likely feature in order to connect the early quantum gravity regime with today observed universe \cite{faraoni}.

Recently, a new cosmological model describing the different stages of evolution of the universe has been proposed \cite{lim}. This model, named Dust of Dark energy, describes  the accelerated expansion of the Universe by two  scalar fields where one of them is a Lagrange multiplier (LM)  which imposes  a constraint on the dynamics \cite{Odin21, LM1, Cap1}.
The extension to $f(R)$ gravity via the addition of a Lagrange multiplier 
constraint has been  proposed in Ref.~\cite{Odin21}. 
Such a model  can be considered as a 
new version of modified gravity because  dynamics and cosmological solutions 
are different from the standard version of $f(R)$ gravity without such a constraint. However, in this article examined the Lagrange multipliers connected with the additional scalar field, while in our work, no additional fields. Furthermore, we consider $f(R, G)$ gravity, not only $f(R)$ model. This result is clear from a dynamical viewpoint: LMs are anholonomic constraints capable of reducing dynamics \citep{arnold,cimento}.
Furthermore, using the LM  approach   helps in the formulation 
of covariant renormalizable gravity \cite{O} and  can be  re-conducted to the existence of Noether  
symmetries into dynamics \cite{annalen,Cap1}.

In the present paper, we study the Gauss-Bonnet gravity with LM
constraints in view to recover realistic dark energy behaviors. Here we do not specify the form of the gravitational action but generically require a function of the Ricci scalar $R$ and the Gauss-Bonnet invariant  $G$, by considering a generic $F(R,G)$-gravity. In this way, due to the form of the Gauss-Bonnet invariant,  all the curvature invariants are taken into account without arbitrary choices. Unlike the models dealt in  \cite{Cap1}, 
where the introduction of LM helps to generate a number of new solutions, we 
found the opposite effect in the case under consideration. Namely, we see 
that the appearance of LMs reduces the number of solutions and does not 
help in their generation. This feature is strictly related to the indetermination in the $F(R,G)$ function endowed with all the gravitational degrees of freedom apart from the standard one in the Hilbert-Einstein action $R$.  The paper is organized as follows. In Sect. II, we  discuss the main features of $F(R,G)$-gravity where a LM is considered for the Gauss-Bonnet component.  In view  of studying dark energy models, the cosmological equations are derived. In particular, we recast the further gravitational degrees of freedom in the form of an effective fluid in order to derive a suitable EoS. Sect.III is devoted to the search for cosmological power law solutions  for some classes of the $F(R,G)$ function. Results are discussed in Sect.IV where we draw also our conclusions.

\section{ $F(R,G)$ gravity  with Gauss-Bonnet  Lagrange multiplier}
As  discussed in \cite{review1,GB11,Cap1}, higher-order and  non local curvature invariants come out in any effective theory derived  from unified approaches like strings, braneworld, etc.
Here, we will consider a generic $F(R,G)$ function of the Ricci scalar $R$ and the Gauss-Bonnet invariant $G$, where we ask for a Lagrange multiplier for the Gauss-Bonnet term in order to constrain cosmology.
The starting action has the following form:
\be
\label{LagS1}
S = \frac{1}{2\kappa^2}\int d^4 x \sqrt{-g} \left\{
F(R,G)+
 \lambda \left( \frac{1}{2} \nabla_\mu G \nabla^\mu G +
F_2(G) \right)\right\}+S_m \, .
\ee
Here $1/2 \kappa^2$ is the gravitational coupling constant, $F(R,G)$ and $F_2(G)$ are arbitrary functions, $S_m$  is the  standard perfect flud matter action, $\lambda$ is a Lagrange
multiplier  whose variation  yields the  constraint equation 
\be
\label{LM_EQ}	
\frac{1}{2}\nabla_\mu G \nabla^\mu G+F_2(G)=0\,.
\ee
 $G$ is the Gauss-Bonnet invariant defined as 
\be
G=R_{\mu\nu\alpha\beta}R^{\mu\nu\alpha\beta}-4R_{\mu\nu}R^{\mu\nu}+R^2\,.
\ee
The gravitational field equations are
\begin{equation}\label{gen-eq}\begin{array}{l}
-\frac{1}{2}Fg_{\mu\nu}+\frac{\lambda}{2}\nabla_\mu G\nabla_\nu G+F'_R R_{\mu\nu}-\nabla_\mu\nabla_\nu F'_R+
g_{\mu\nu}\Box
F'_R+2FL'_{G}\(R_\mu^{\phantom{\mu}\alpha\beta\gamma}R_{\nu\alpha\beta\gamma}+
2R_{\mu\alpha\beta\nu}R^{\alpha\beta}-\right.\vphantom{\pi
G{c^4}}\\
\quad{}\left.-2R_{\alpha\mu}R^\alpha_{\phantom{\alpha}\nu}+RR_{\mu\nu}\)
-4\(R_{\mu\alpha\beta\nu}+R_{\mu\nu}g_{\alpha\beta}+R_{\alpha\beta}g_{\mu\nu}-
R_{\alpha\mu}g_{\beta\nu}-R_{\alpha\nu}g_{\beta\mu}+\right.\vphantom{8\pi
G{c^4}}\\
\quad{}\left.+\frac{1}{2}R\(g_{\alpha\mu}g_{\beta\nu}-g_{\mu\nu}
g_{\alpha\beta}\)\)\nabla^\alpha\nabla^\beta
FL'_G=\kappa^2 T_{\mu\nu}.\end{array}\end{equation}
Here $ F'_R=dF(R,G)/dR$ and $FL'_G=dF(R,G)/dG+\lambda dF_2(G)/dG-\nabla^\mu (\lambda \nabla_\mu G)$.
In  the 4-dimensional space-time, we have the following expression
$$2\(R_\mu^{\phantom{\mu}\alpha\beta\gamma}R_{\nu\alpha\beta\gamma}+
2R_{\mu\alpha\beta\nu}R^{\alpha\beta}-2R_{\alpha\mu}R^\alpha_{\phantom{\alpha}\nu}+RR_{\mu\nu}\)-
\dac{1}{2}{G}g_{\mu\nu}=0.$$ 
Hence Eq. (\ref{gen-eq}) can be rewritten as
\begin{equation}\label{eq-4D}\begin{array}{l}
-\frac{1}{2}Fg_{\mu\nu}+\frac{\lambda}{2}\nabla_\mu G\nabla_\nu G+F'_R R_{\mu\nu}-\nabla_\mu\nabla_\nu F'_R+
g_{\mu\nu}\Box F'_R+\frac{1}{2}FL'_{G}{G}g_{\mu\nu}
-4\(R_{\mu\alpha\beta\nu}+R_{\mu\nu}g_{\alpha\beta}+\right.\vphantom{\dac{8\pi
G}{c^4}}\\ \quad{}\left.+R_{\alpha\beta}g_{\mu\nu}-
R_{\alpha\mu}g_{\beta\nu}-R_{\alpha\nu}g_{\beta\mu}+\frac{1}{2}R\(g_{\alpha\mu}g_{\beta\nu}-g_{\mu\nu}
g_{\alpha\beta}\)\)\nabla^\alpha\nabla^\beta
FL'_{G}=\kappa^2 T_{\mu\nu}\,.\end{array}\end{equation}

Let us now consider  a Friedmann-Robertson-Walker (FRW)  universe with the
flat spatial metric
\be
\label{Lag7} ds^2 = - dt^2 + a(t)^2 \sum_{i=1,2,3}
\left(dx^i\right)^2\, .
\ee
Then from Eq. (\ref{LM_EQ}) one  obtains
\be
\label{LagS2} F_2(G) = \frac{{\dot G}^2}{2} \; .
\ee
FRW equations are specified as soon as one defines the effective density and pressure  
\begin{eqnarray}
&& \rho_{\mathrm{eff}} \equiv 
\frac{1}{{{F'}}_{R}} \left\{ \rho_{\mathrm{M}} +
\frac{1}{2\kappa^{2}}
\left[ \left( {{F'}}_{R}R+G\;{{FL'}}_{G}-{F} \right)-\lambda\dot{G}^2
-6H{\dot{{F'}}}_{R}
-24H^3{\dot{{FL'}}}_{G}
\right] \right\}\,,
\label{LagS3:1} \\
&& 
p_{\mathrm{eff}} \equiv 
\frac{1}{{{F'}}_{R}} \left\{p_{\mathrm{M}} +
\frac{1}{2\kappa^{2}} \left[
-\left( {{F'}}_{R}R+G\;{{FL'}}_{G}-{F} \right)
+4H{\dot{{F'}}}_{R}+2{\ddot{{F'}}}_{R}
\right. 
\right. \nonumber \\ 
&& \left. \left. 
{}
+16H\left(\dot{H} +H^2 \right){\dot{{FL'}}}_{G}
+8H^2 {\ddot{{FL'}}}_{G}
\right]
\right\}\;.
\label{LagS3:2}
\end{eqnarray}
Here
$\rho_{\mathrm{eff}}=3\kappa^{-2}H^{2}$ 
and 
$
p_{\mathrm{eff}}=-\kappa^{-2} \left( 2\dot H+3H^{2} \right)
$, respectively. 
A general method for  solving such equations does not exist.  However,  a few solutions for general $F (R, G)$ theory have been found \cite{Nojiri:2005sx, Bamba:2009uf,AM,mak11, Bamba:2013}.
We can reduce the problem,  if we assume that
\be
FL'_G=F'_G,\;\;\lambda\dot{G}^2=2\kappa^2\rho_{\mathrm{M}},\;\;p_{\mathrm{M}}=0.
\ee
We obtain an equation for the Lagrange multiplier,  which will lead us to a model where the material field is in the form of dust ($p=0$).   In such a case 
the conservation equation for matter
\be \dot{\rho_{\mathrm{M}}}+3 H(\rho_{\mathrm{M}}+p_{\mathrm{M}})=0\,,
\ee
could result violated but the general Bianchi identities have to remain conserved.
Let us consider in more detail Eqs. (\ref{LagS3:1})-(\ref{LagS3:2}). It is easy to remove from these  equations the constraint derived from the Lagrangian multiplier. We obtain then  the compatibility condition of the equations, which takes the form

\bea
&&3 H \kappa^2 p_{\mathrm{M}}+3 H\kappa^2 \rho_{\mathrm{M}}+\kappa^2 \dot{\rho_{\mathrm{M}}}+
288 H\lambda \left[2 \dot H \left(2 H^2+ \dot H\right)+H  \ddot H \right] \times\\
\times&&\left[2 \left( \dot H^2 \left(-6 H^2+96 H^6+ \dot H(-1+96 H^4+24 H^2 \dot H)\right)+\right.\right. \nonumber\\
  +&&H\left.\left. \left(-2 H^2+3  \dot H (-1+16 H^4+8 H^2  \dot H)\right) \ddot H+6 H^4  \ddot H^2\right)-H^2 \dddot{ H}\right]=0\nonumber.
\eea
This equation can be rewritten in terms of the Gauss-Bonnet invariant obtaining 
\be
\label{eq_S}
\lambda \dot G\left(\dot G^2-2\ddot G\right)=-4\kappa^2\left(3 H ( p_{\mathrm{M}}+\rho_{\mathrm{M}})+\dot{\rho_{\mathrm{M}}}\right).
\ee
If the matter satisfies the conservation law we need to impose the additional conditions
$ G = \mbox{const}$ or  $G = c_1 +2 \ln\left(t + c_2\right)$.

A similar situation arises when we consider another  Lagrangian  of slightly different type, that is 
\be
\label{LagS11}
S = \frac{1}{2\kappa^2}\int d^4 x \sqrt{-g} \left\{
F(R,G)+
 \lambda \left( \frac{1}{2} \nabla_\mu R \nabla^\mu R +
F_2(R) \right)\right\}+S_m \, ,
\ee
where $\lambda$ is again a Lagrange multiplier and $F_2$ is an arbitrary function of the scalar curvature $R$.

In this case the compatibility condition of the equations of motion will have exactly the same form, but with the replacement of the invariant of the Gauss-Bonnet  with  the Ricci  curvature scalar 
\be
\label{eq_S1}
\lambda \dot R\left(\dot R^2-2\ddot R\right)=-4\kappa^2\left(3 H ( p_{\mathrm{M}}+\rho_{\mathrm{M}})+\dot{\rho_{\mathrm{M}}}\right).
\ee
Clearly Eqs. (\ref{eq_S}) and (\ref{eq_S1}) do not contradict the validity of the general Bianchi identities for the theory but point out that (Gauss-Bonnet) curvature terms, selected by the Lagrange multipliers, can act as source terms.  
With these considerations in mind, let us search for cosmological solutions for $F(R,G)$ gravity according to the Lagrange multiplier method.

\section{Cosmological solutions for $F(R,G)$ gravity}
Our aim is now to search for cosmological solutions compatible with $F(R,G)$ gravity that, eventually, select the form of the function. Let us start with power law solutions where 
 the scale factor behaves as
\be
\label{power}
a(t)=a_0 t^{h_0},
\ee
so that the Hubble rate is given by
\be
H=\frac{h_0}{t}.
\ee
The FRW equations assume the form
\bea
0&=&(F-2\kappa^2\rho_{\mathrm{M}})  t^{14}-24 (h_0-1) {h_0}^3 t^{10}F(R,G)'_G-6 (h_0-1) {h_0} t^{12}F(R,G)'_R+\nonumber\\
&+& 6 {h_0} \left(-384(h_0-1) {h_0}^5 \left(48(h_0-1) {h_0}^3 (9+{h_0})+(29-{h_0} (14+3 {h_0})) t^4\right)\lambda+\right.\\
&+&\left.  \left(4 {h_0}^2 t^{10} \dot{F(R,G)'_G}+t^{12}\dot{F(R,G)'_R}+384(h_0-1) {h_0}^5 \left(\left(48 (h_0-1) {h_0}^3+(9-2 {h_0}) t^4\right) \dot{\lambda}-t^5 \ddot{\lambda}\right)\right)\right),\nonumber
\eea
and
\bea
0&=&(F+2\kappa^2p_{\mathrm{M}}) t^{14}-24(h_0-1) {h_0}^3 t^{10}F(R,G)'_G+2 (1-3 {h_0}) {h_0} t^{12} F(R,G)'_R+\nonumber\\
&+&2 \left(-1152 (h_0-1) {h_0}^5 \left(16(h_0-1) {h_0}^3 (10+{h_0})(3h_0-13)-(h_0-3) (6+{h_0}) (3h_0-5) t^4\right) \lambda+\right.\\
&+& t \left(8 (h_0-1) {h_0}^2 t^{10}\dot{F(R,G)'_G}+2 {h_0} t^{12} \dot{F(R,G)'_R}-405504 {h_0}^8 \dot{\lambda}+847872 {h_0}^9 \dot{\lambda}-479232 {h_0}^{10} \dot{\lambda}+36864 {h_0}^{11} \dot{\lambda}+\right.\nonumber\\
&+&42240 {h_0}^5 t^4 \dot{\lambda}-64896 {h_0}^6 t^4 \dot{\lambda}+23808 {h_0}^7 t^4 \dot{\lambda}-1152 {h_0}^8 t^4 \dot{\lambda}+4 {h_0}^2 t^{11}\ddot{F(R,G)'_G}+t^{13}\ddot{F(R,G)'_R}+18432 {h_0}^8 t \ddot{\lambda}-\nonumber\\
&-&\left.\left.36864 {h_0}^9 t \ddot{\lambda}+18432 {h_0}^{10} t \ddot{\lambda}-6528 {h_0}^5 t^5 \ddot{\lambda}+8448 {h_0}^6 t^5 \ddot{\lambda}-1920 {h_0}^7 t^5 \ddot{\lambda}-384 (h_0-1) {h_0}^5 t^6 {\lambda}^{(3)}\right)\right).\nonumber
\eea
If we choose the Lagrange multiplier as
\be
\label{LM_1}
\lambda= c\; e^{-\frac{12(h_0-1) {h_0}^3}{t^4}} t^{5-3 h_0},
\ee
were $c$ is a constant, matter behaves as "dust"  assuming the form
\be 
\label{LM_M}
\rho_{\mathrm{M}}=\frac{9216 c}{2\kappa^2}  e^{-\frac{12(h_0-1) {h_0}^3}{t^4}}(h_0-1)^2 {h_0}^6 t^{-5-3{h_0}} ,\;\;\; p_{\mathrm{M}}=0\,,
\ee
where  the pressure is clearly zero.

Choosing, instead, a Lagrange multiplier of the form 
\be
\label{LM_Ex}
\lambda=\lambda_1 t^{\lambda_2}\,,
\ee
one can use the reconstruction method  proposed in Refs. \cite{Cap1,Nojiri:2005sx, Bamba:2009uf, Bamba:2013}.  However, in this case we need to introduce the matter in the special form
\bea
\label{LM_Ex1}
\rho_{\mathrm{M}}&=&c_1 t^{-3h_0(1+w)}+\frac{4608(h_0-1)^2 h_0^6 \lambda_1 t^{-14+\lambda_2}}{\kappa^2}  \left(\frac{48(h_0-1){h_0}^3}{-14+\lambda_2+3 h_0 (1+w)}-\frac{5 t^4}{-10+\lambda_2+3 h_0 (1+w)}\right),\\
p_{\mathrm{M}}&=&w \rho_{\mathrm{M}}.
\eea
If $\lambda_2=14-3h_0(1+w)$ then the matter density takes the form

\be
\label{LM_Ex2}
\rho_{\mathrm{M}}=t^{-3 h_0 (1+w)} \left(c_1+\frac{1152 (h_0-1)^2h_0^6 \lambda_1\left(-5 t^4+192 (h_0-1)h_0^3 \ln{t}\right)}{\kappa^2}\right)\,.
\ee
On the other hand,  if $\lambda_2=10-3h_0(1+w)$,  the matter density takes the form
\be
\label{LM_Ex3}
\rho_{\mathrm{M}}=t^{-3 h_0 (1+w)} \left(c_1+\frac{4608(h_0-1)^2 h_0^6 \lambda_1 \left(-\frac{12(h_0-1)h_0^3}{t^4}-5\ln{t}\right)}{\kappa^2}\right)\,.
\ee

With proper functions $P(\phi)$, $Z(\phi)$ and $Q(\phi)$ of a
scalar field $\phi$, which we can identify with the time $t$, 
we represent the term $\mathcal{F}(R,G)$ in the action in Eq.~({\ref{LagS1}}) 
as $P(t)R+Z(t)G+Q(t)$. 
Varying this action with respect to $t$, we find 
$\left( dP(t)/dt \right)R + \left( dZ(t)/dt \right)G + dQ(t)/dt = 0$. 
By solving this equation, we obtain $t=t(R,G)$. 
Combining this and using the above representation, 
we get the special case  $\mathcal{F}(R,G) = P(t)R+Z(t)G+Q(t)$. 

Let us consider now a few  cases that can be worked out in detail.

\subsection{The case  of $F(G)$ gravity with a Lagrange multiplier}
Let us start with a function  the form  $F=R+F(G)$, where we consider a pure Gauss-Bonnet function as correction to Einstein gravity.
 Considering $c_1=0$ (without  ordinary matter) then it is easy to find an explicit form for  $F(G)$.
There are two cases
\begin {enumerate}
\item
We have 
\bea
\label{Ex_3}
F'_G&=&\frac{4(h_0-1) h_0^2 (3h_0-2) \left(-10+h_0+3h_0^2\right)}{(h_0-2)^2 (5+h_0) \left(-1+2h_0+3h_0^2\right) t^2},\nonumber\\
\lambda&=&\frac{(2-3h_0) t^8}{1152 h_0^4 \left(-2+7 h_0-h_0^2-7 h_0^3+3h_0^4\right)},\nonumber\\
t&\to& \frac{2^{3/4} 3^{1/4} \left(-h_0^3+h_0^4\right)^{1/4}}{G^{1/4}},\\
F(G)&=&\frac{2 \sqrt{\frac{2}{3}} G^{3/2} \sqrt{(h_0-1) h_0^3} \left(20-32 h_0-3 h_0^2+9h_0^3\right)}{3 (h_0-2)^2 h_0 (5+h_0) \left(-1+2 h_0+3 h_0^2\right)},\nonumber
\eea
where the Gauss-Bonnet function scales as a $3/2$ power. The dark energy behavior means that the power $h_0$ in Eq.(\ref{power}) is larger than 1.

\item
Another case is 

\bea
\label{Ex_4}
F'_G=\frac{\left(84-20 h_0-143 \text{h0}^2-30h_0^3+9 h_0^4\right) t^6}{192 (-3+h_0) (-1+h_0) h_0^4 (2+3h_0) \left(2-7 h_0+3 h_0^2\right)}\nonumber,\\
\lambda=\frac{(-2+3 h_0) t^{12}}{18432 (-1+h_0)^2 h_0^7 \left(2-7 h_0+3 h_0^2\right)},\\
F=\frac{\sqrt{\frac{3}{2}} \sqrt{(-1+h_0) h_0^3} \left(-84+20h_0+143 h_0^2+30 h_0^3-9h_0^4\right)}{\sqrt{G} (-3+h_0) h_0 (2+3 h_0) \left(2-7 h_0+3 h_0^2\right)}.\nonumber
\eea
If we eliminate the Lagrange multiplier we get the following form of the function $ F (G)$

\bea
\label{Ex_5}
F'_G&=&g_1 t^{11/3},\nonumber\\
F(G)&=&\frac{32}{27}g_1 \left(\sqrt{2}\; 3^{1/4}+11 G^{1/4}\right) G^{1/12}
\eea
and then we get another fractional power for $F(G)$. It is worth noticing that this kind of fractional behavior for generic gravitational actions  is usually determined by the presence of Noether symmetries for the dynamical system (see \cite{annalen} for details). This shows the strict relation between LM method and the Noether Symmetry Approach to reduce dynamical systems. 

\end{enumerate}

\subsection{The case of $F(R,G)$ gravity in presence of ordinary matter}

Let us assume now a generic $F(R,G)$ function with no Lagrange multiplier, but the presence of matter in the form of a standard perfect fluid like 
$\rho_{\mathrm{M}} =\rho_0 a^{-3(1+w)}$, $p_{\mathrm{M}}= w \rho_0 a^{-3(1+w)}$. The FRW Eqs. (\ref{LagS3:1}-\ref{LagS3:2}) take the following form

\bea
-2\kappa^2 \rho_0 t^4 + 
 a_0^3 t^{3 h_0} (a_0 t^{h_0})^{ 3 w} (F t^4 - 
    6 h_0 (4g_1 h_0^2 (-1 -g_2 + h_0) t^{f_2} + 
       r_1 (-1 + h_0 - r_2) t^{2 + r_2}))=0,
\label{FRG11}
\eea

\bea
{a_0}^3 t^{3 h_0} (a_0 t^{h_0})^{ 3 w} \left(-F t^4 - 8 g_1 (g_2 - h_0) h_0^2 (-3 + f_2 + 3 h_0) t^{f_2} + 
    2 r_1 (h_0 - r_2) (-1 + 3 h_0 + r_2) t^{2 + r_2}\right) &-& \nonumber\\
-2 \kappa^2 \rho_0 t^4 w&=&0.
\label{FRG12}
\eea

Here, the functions $F'_R$ and $F'_G$ can be  selected in the form $F'_R=r_1 t^{r_2}$ and $F'_G=g_1 t^{g_2}$.  Using the reconstruction method \cite{Nojiri:2005sx, Bamba:2009uf, Bamba:2013} which we described above (that is $Z(t)\equiv F'_G$, $P(t)\equiv F'_R$ and $Q(t)=F(R,G)- F'_R R-F'_G G$), 
we can find $t$ as a function of $R$ and $G$, and then $F$ as a function of $R$ and $G$. However, this will lead to an equation containing  an  irrational degree for $t$, which cannot be solved in general. As above, one can consider a few special cases, but just for fixed values of $ h_0 $ and $ w $. In fact, deriving  the explicit form of  $F$ as a function of $R$ and $G$ is very difficult.
We choose, for the sake of simplicity, the case when $ F (R,G)=F_1(R)+F_2(G)$. Then the solution is possible, for $\rho_0\ne0$, if 

\begin{enumerate}

\item  $ g_2=4-3h_0(1+w) \ne 2+r_2$\;,

\item   $2+r_2=4-3h_0(1+w) \ne g_2$\;,

\item  $g_2=2+r_2=4-3h_0(1+w)$\;.
\end{enumerate}
Let us  consider the first case. Solving Eqs. (\ref{FRG11})-(\ref{FRG12})  we find
\bea
\rho_0&=&-4 a_0^{3 + 3 w} g_1 h_0^2 (-4 + 3 h_0 (1 + w) (-1 + h_0 (4 + 3 w))/(\kappa^2 (1 + w)),\\
r_2&=& \frac{1}{2} \left(1+h_0 \pm \sqrt{1+10h_0+h_0^2}\right).
\eea
It is evident that, in order that the condition $\rho_0> 0$ be satisfied,   we need to impose restrictions 
on $g_1$ and $h_0$, namely:  $g_1>0$ and $1/4<h_0<4/3$  or  $g_1>0$ and $h_0<1/4$ or $h_0>4/3$. 
Finally the functions $F_1(R)$  and  $F_2(G)$  assume the form

\bea
\label{Ex_6}
F_1(R)=\frac{2^{\frac{1}{4} \left(9+h_0\pm\sqrt{1+10 h_0+h_0^2}\right)} 3^{\frac{1}{4} \left(1+h_0\pm\sqrt{1+10 h_0+h_0^2}\right)} (h_0 (2 h_0-1))^{\frac{1}{4} \left(1+h_0\pm\sqrt{1+10 h_0+h_0^2}\right)} R^{\frac{1}{4} \left(3-h_0\mp\sqrt{1+10 h_0+h_0^2}\right)} {r_1}}{3-h_0\mp\sqrt{1+10 h_0+h_0^2}}
\eea

\bea
\label{f1}
F_2(G)=g_1\frac{ 2^{5 - \frac{9}{4} h_0 (1 + w)} 3^{-\frac{3}{4} h_0 (1 + w)}(h_0-1)^{1 -\frac{3}{4} h_0 (1 + w)} h_0^{ 2 -\frac{39}{4} h_0 (1 + w)}}{1 + w} G^{
 \frac{3}{4} h_0 (1 + w)} .
\eea

In  the second case the functions $F_{1,2}$ assume  the forms
\bea
\label{f2}
F_1(R)=\frac{{r_1}2^{2-\frac{3}{2}h_0 (1+w)} 3^{-\frac{3}{2}h_0 (1+w)}(2h_0-1) (h_0(2h_0-1))^{-\frac{3}{2}h_0 (1+w)} }{1+w}  R^{\frac{3}{2} h_0 (1+w)},
\eea

\bea
\label{Ex_7}
F_2(G)=-{g_1}2^{2+\frac{3 (3+h_0)}{4}} 3^{\frac{3+h_0}{4}}  (h_0-1)^{\frac{1}{4} (h_0-1)} {h_0}^{\frac{3 (3+h_0)}{4}}G^{\frac{1}{4}-\frac{h_0}{4}} .
\eea
If $\rho_0 = 0$ then the function $F$ is the sum of the functions $F_2 (G)$ of the first case (\ref{f1}) and the functions $F_1 (R)$ of  the second case (\ref{f2}).

For the third case we  easily find the function $F$ by summing up the functions

\bea
\label{Ex_8}
F_1(R)= {r_1}\frac{2^{2-\frac{3}{2} {h_0} (1+w)} 3^{-\frac{3}{2} {h_0} (1+w)}(2h_0-1) ({h_0} (-1+2 {h_0}))^{-\frac{3}{2} {h_0} (1+w)}}{1+w} R^{\frac{3}{2} {h_0} (1+w)},
\eea

\bea
\label{Ex_9}
F_2(G)={g_1}\frac{2^{5-\frac{9}{4} {h_0} (1+w)} 3^{-\frac{3}{4} {h_0} (1+w)}  (h_0-1)^{1-\frac{3}{4} {h_0} (1+w)} {h_0}^{2-\frac{9}{4} {h_0} (1+w)}}{1+w}G^{\frac{3}{4} {h_0} (1+w)} .
\eea
As general remark, also in these cases the functions $F$ are fractional power laws pointing out that their form could be easily related to the presence of a Noether symmetry for the dynamical system.

\section{Discussion and conclusions}
We have examined  cosmological models derived  from  $ F (R, G)$ gravity in presence of   LMs like in the action  (\ref{LagS1}).  No other  additional field has been considered  with the exception of ordinary perfect fluid matter.
Despite of Refs. \cite{LM1, Cap1}, where with the help of  LMs  new solutions were generated, here, we have constraints  like  (\ref{eq_S})  and (\ref{eq_S1}) that have to be  taken into account  restricting the possible forms of the  function $F(R,G)$. On the other hand  we  can   introduce,  in addition to  standard  matter (in the form of a perfect fluid satisfying the conservation equation),  some form of  exotic matter, which should  compensate,  in Eqs. (\ref{eq_S})  and  (\ref{eq_S1}), the  terms corresponding to LMs.
For  a power law behavior of the cosmological scale factor $a(t)$  and the LM of  the form (\ref{LM_1}),  we have to introduce  matter in the form  (\ref{LM_M}),  for which our  model  reduces to the standard $ F (R, G)$ model without  LMs. 

To generate other solutions one can choose the LM in the form  (\ref{LM_Ex}); then  the matter of the form
(\ref{LM_Ex3}) will provide the compatibility of  FRW  equations.
All of these examples  are obtained for  power law  behaviors of the scale factor.
Another example is based on  $F (G)$ gravity. In this case it is easy to reconstruct the explicit form of the function $F(R)$. For a LM in the form (\ref{LagS1}) this function has the form (\ref{Ex_3}) and (\ref{Ex_4}); with the exclusion of the LM, the function $F (G)$ has the form  (\ref{Ex_5}).
All of these solutions have been built in the absence of ordinary matter, but the addition of such matter does not change  much the situation. As examples, we have built several solutions for  $ F(R)+F(G)$  gravity with  matter, but without  LMs   like (\ref{Ex_5}) -(\ref{Ex_9}). Adding LM, as noted above,  leads to the appearance of exotic matter and significant complication in equations: this means   a series of additional restrictions on the $F(R,G)$ functions.
In conclusion,   the introduction of  LMs can lead to additional restrictions, which can be avoided in the models with further fields, such as scalar fields. In this case, the compatibility of the FRW equations is achieved by the right choice of a scalar field. Standard  choices for it, for example, $\phi = \ln t$,  allow  to give rise to   consistent sets of equations as discussed in \cite{Cap1}. 
Finally, we can say that the LM method is a sort of  inverse scattering approach that allows to reconstruct a model, given the cosmological behavior that one wants to fit. This feature is particularly relevant with respect to the issue of matching models with observational data instead of imposing, a priori,  arbitrary theories.

\section*{Acknowledgments.}
S.C. is supported by INFN Sez. di Napoli (Iniziative Specifiche NA12 and OG51). A.N.M. is supported by projects N 2.1839.2011 and N 14.B37.21.0774 of Min. of Education and Science (Russia) and LRSS project 224.2012.2 (Russia). The Authors acknowledge their friend S.D. Odintsov for useful suggestions and discussion on the topics of this paper.

\end{document}